\documentclass[aps,prb,amsmath,amssymb,
preprint,%
]{revtex4-2}

\usepackage{lmodern}
\usepackage[bookmarks,%
            colorlinks,%
            urlcolor=blue,%
            citecolor=blue,%
            linkcolor=blue,%
            hyperfigures,%
            pdfcreator=KN,%
            breaklinks=false]{hyperref}
\usepackage[capitalise]{cleveref}
\usepackage[english]{babel}
\usepackage{siunitx}
\usepackage{csquotes}
\usepackage{nth}
\usepackage{multirow}
\usepackage[version=4]{mhchem}
\usepackage{graphicx}
\usepackage[inline]{enumitem}

 \usepackage{xr}
\usepackage[caption=false]{subfig}

\DeclareGraphicsExtensions{.pdf}

\begin{document}

\title{History- and Frequency-Dependent Dielectric Nonlinearities Induced by Polar Nanoregions in \ce{0.5(Ba_{0.7}Ca_{0.3}TiO_{3})-0.5(BaZr_{0.2}Ti_{0.8}O_{3})} Thin Films}
\author{Kevin Nadaud}\altaffiliation{Author to whom correspondence should be addressed: \href{mailto:kevin.nadaud@univ-tours.fr}{kevin.nadaud@univ-tours.fr}}
\affiliation{%
GREMAN UMR 7347, Université de Tours, CNRS, INSA-CVL, 16 rue Pierre et Marie Curie, 37071 Tours, France%
}%
\author{Guillaume F. Nataf}
\affiliation{%
GREMAN UMR 7347, Université de Tours, CNRS, INSA-CVL, 16 rue Pierre et Marie Curie, 37071 Tours, France%
}
\author{Nazir Jaber}
\affiliation{%
GREMAN UMR 7347, Université de Tours, CNRS, INSA-CVL, 16 rue Pierre et Marie Curie, 37071 Tours, France%
}
\author{Edgar Chaslin}
\affiliation{%
GREMAN UMR 7347, Université de Tours, CNRS, INSA-CVL, 16 rue Pierre et Marie Curie, 37071 Tours, France%
}

\author{Béatrice Negulescu}
\affiliation{%
GREMAN UMR 7347, Université de Tours, CNRS, INSA-CVL, 16 rue Pierre et Marie Curie, 37071 Tours, France%
}
\author{Jérôme Wolfman}
\affiliation{%
GREMAN UMR 7347, Université de Tours, CNRS, INSA-CVL, 16 rue Pierre et Marie Curie, 37071 Tours, France%
}


\begin{abstract}
    In this article, dielectric non-linearities in \ce{0.5(Ba_{0.7}Ca_{0.3}TiO_{3})-0.5(BaZr_{0.2}Ti_{0.8}O_{3})} thin film are studied using impedance spectroscopy and harmonic measurements, as a function of the AC measuring field, at different frequencies and upon cycling.
    The measurements reveal that the pinching of the hysteresis loop, characterized by a phase angle of the third harmonic close to \qty{-270}{\degree}, is stronger for low frequencies.
    This confirms that the pinching is induced by the presence of polar nanoregions (PNRs), whose responses are also strongly dependent on frequency.
    When repeating the measurement, the PNR contribution changes since the phase angle of the third harmonic response evolves from pinched to conventional relaxor.
    This shows that strong changes in the PNR configuration can be induced, even for low AC fields.
    First Order Reversal Curves (FORC) confirm the presence of a pinched hysteresis loop.
    When repeating the FORC measurement a second time, the distribution drastically changes and corresponds to a soft ferroelectric.
    The asymmetry of the Preisach plane measured using FORC is confirmed by a proposed novel measurement strategy: unipolar impedance measurements.
\end{abstract}
\maketitle
\section{Introduction}

The high dielectric permittivity and piezoelectric coefficient of relaxor ferroelectrics make them promising materials for energy storage and actuator applications~\cite{JayakrishnanPMS2023,VeerapandiyanM2020,PalneediAFM2018,PrateekCR2016}.
Relaxors exhibit polar nanoregions (PNRs), clusters in which the polarization is randomly aligned in absence of external electric field~\cite{CollaJAP1998,CowleyAP2011,BokovJMS2006,OtonicarAFM2020}.
For relaxors and conventional ferroelectrics, the polar cluster boundaries, PNRs and ferroelectric domain walls (DWs) respectively, contribute significantly to the macroscopic dielectric and piezoelectric responses ~\cite{Li2016,LiAFM2018,LiAFM2017,DamjanovicJACS2005}.
Those PNRs can subsist several kelvins above the Curie temperature (in the centrosymmetric phase) ~\cite{GartenJACS2016,GartenJAP2014,SaljePRB2013}, which results in a dielectric permittivity peak much broader in temperature than that of conventional ferroelectric ~\cite{CowleyAP2011} and frequency dependent~\cite{DecFerroelectrics2008,GlazounovPRB2000,DecPRB2003}.

Due to their non-linear response to the applied electric field, ferroelectrics exhibit specific dielectric responses, even below the coercive field.
The measurement of the dielectric permittivity as function of the AC field can be used to determine the reversible and irreversible domain wall motion contributions.
According to the Rayleigh law, the relative permittivity linearly increases:~\cite{taylorjap1997,damjanovicrpp1998}
\begin{equation}
    \varepsilon_{r} = \varepsilon_{\mathit{r-l}} + \alpha_{r}E_{\mathit{AC}}
    \label{rayleigh}
\end{equation}
$\varepsilon_{\mathit{r-l}}$ correspond to the low field permittivity and $\alpha_{r}$ to the slope and represents the irreversible domain wall motion contribution.
The Rayleigh law assumes that domain walls interact with a homogeneous distribution of the pinning centers.

However, in real ferroelectric materials, the distribution of pinning centers is not homogeneous and the linear increase is only visible after a given threshold~\cite{HallF1999,SchenkPRA2018,TaylorAPL1998}.
For low field, domain walls can only vibrate around an equilibrium position and the contribution is called reversible.
In that case, the relative permittivity evolution with the measuring field can be described using the hyperbolic law:~\cite{NadaudAPL2024,borderonapl2011,BaiCI2017}
\begin{equation}
    \varepsilon_{r} = \varepsilon_{\mathit{r-l}} + \sqrt{\varepsilon_{\mathit{r-rev}}^2 + (\alpha_{r}E_{\mathit{AC}})^2}
    \label{hyperbolic}
\end{equation}
with $\varepsilon_{\mathit{r-rev}}$ the reversible domain wall motion contribution, proportional to the domain wall density~\cite{boserjap1987,NadaudAPL2022,BorderonSR2017}.

In addition to the conventional impedance spectroscopy, which considers only the first harmonic of the polarization, it is possible to study higher order harmonics.
When the applied electric field is $E(t) = E_{0}\sin\left(\omega t\right)$, the polarization can be expressed as :~\cite{RiemerJAP2021}
\begin{align}
    \label{eq:polarization fourier}
    P(t) &= P^{(0)} + \sum_{n= 1}^{\infty}P^{(n)}\sin\left(n\omega t + \delta_{n}\right) \\
         &= P^{(0)} + \sum_{n= 1}^{\infty}P'^{(n)}\sin\left(n\omega t\right) +P''^{(n)}\cos\left(n\omega t\right) 
\end{align}
$P^{(0)}$ is an eventual DC contribution, $P^{(n)}$ is the magnitude of the polarization of the $n$\textsuperscript{th} order and $\delta_{n}$ is its phase.
$P'^{(n)}$ and $P''^{(n)}$ represent the in-phase and quadrature component of the polarization of the $n$\textsuperscript{th} order.
Depending on the analysis, either magnitude/phase or in-phase/quadrature representations are used.
With that decomposition, we can define the $n$\textsuperscript{th} order relative real and imaginary part of the permittivity:
\begin{equation}
    \varepsilon_{r}'^{(n)} = \frac{P'^{(n)}}{E_{0}} \quad\text{and}\quad \varepsilon_{r}''^{(n)} = \frac{P''^{(n)}}{E_{0}}
\end{equation}

\begin{figure}
    \centering
    {\includegraphics[width=0.45\textwidth]{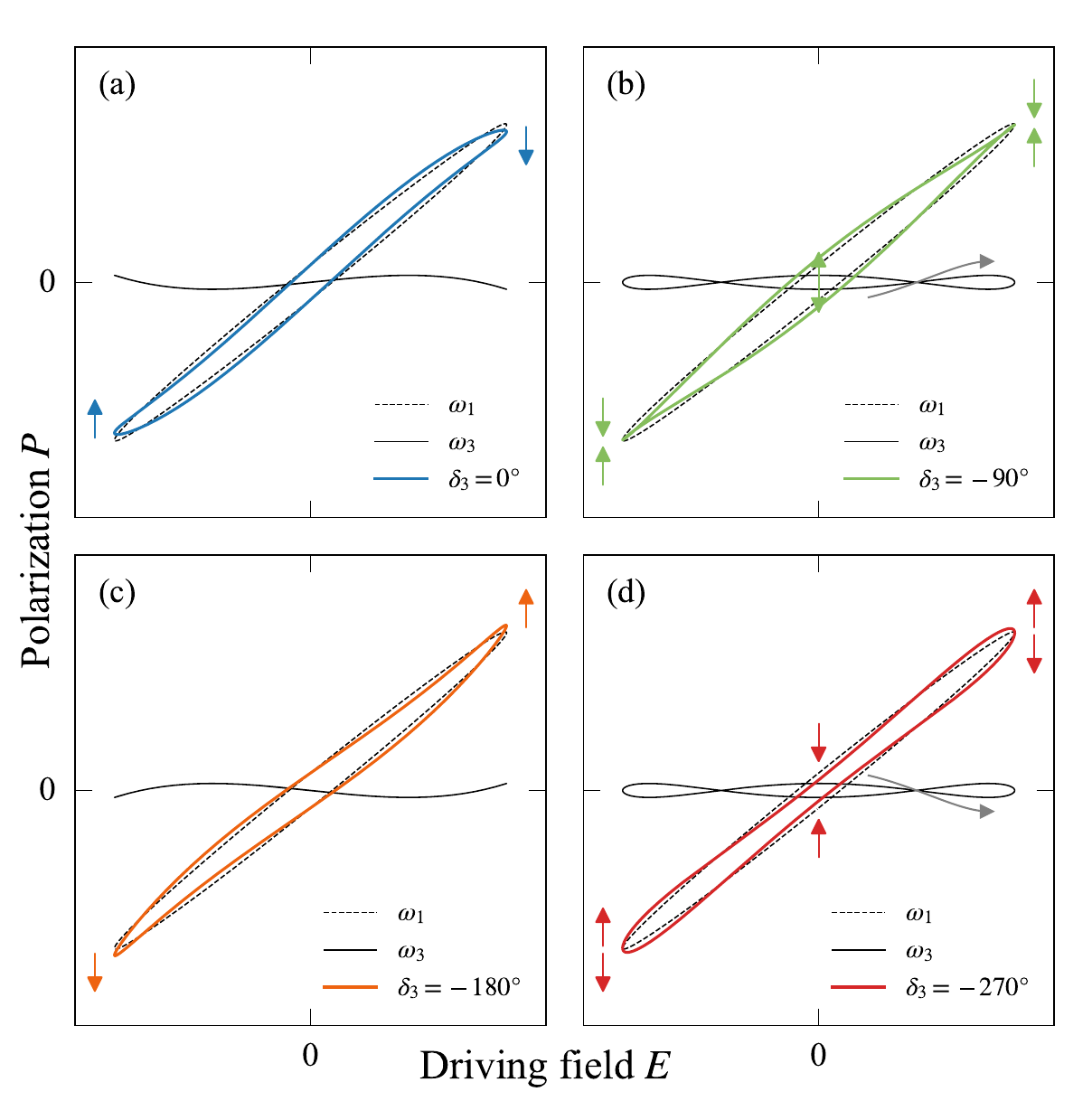}
    \subfloat{\label{subfig:phasor 0}}%
    \subfloat{\label{subfig:phasor -90}}%
    \subfloat{\label{subfig:phasor -180}}%
    \subfloat{\label{subfig:phasor -270}}%
    }
    \caption{Hysteresis deformation due to third-harmonic response shown for four limiting cases: $\delta_{3} = \qty{0}{\degree}$ (a), $\delta_{3} = \qty{-90}{\degree}$ (b), $\delta_{3} = \qty{-180}{\degree}$ (c) and $\delta_{3} = \qty{-270}{\degree}$ (d). Dashed curves correspond to the contribution of the first harmonic and solid black curve to the contribution of the third harmonic. Gray arrows indicate the rotational sense of the third harmonic and colored arrows the deformation of the hysteresis. Adapted from ~\cite{RiemerJAP2021,OtonicarJACS2022}.}
    \label{fig:phasor}
\end{figure}

For ferroelectrics, the phase angle of the third harmonics is particularly studied~\cite{MorozovJAP2008,OtonicarJACS2022,NadaudAPL2024}.
For an ideal ferroelectric, having domain walls that interact with a homogeneous distribution of pinning centers, a phase angle of the third harmonic $\delta_{3} =\qty{-90}{\degree}$ is found as stated by the Rayleigh law~\cite{HashemizadehAPL2017}.
For many ferroelectric materials, the phase angle of the third harmonic differs, at least in some range of AC field, from the ideal value of \qty{-90}{\degree}.
A deviation from this angle gives information regarding the deformation of the polarization loop.
\begin{enumerate}[label=(\roman*)]
    \item $\delta_{3} = \qty{0}{\degree}$, the third harmonic is of opposite sign with respect to the first harmonic, when the field is maximum. This corresponds to saturation-like response (Fig.~\ref{subfig:phasor 0}). A saturation like response is found at large electric field for ferroelectrics and relaxors.
    \item $\delta_{3} = \qty{-90}{\degree}$, the third harmonic opens the center of the loop. This corresponds to the ideal Rayleigh behavior (Fig.~\ref{subfig:phasor -90}), which is found for \ce{BaTiO3}~\cite{HashemizadehAPL2017} and weakly doped \ce{(Pb, Zr)TiO3} ~\cite{MorozovJECS2005}.
    \item $\delta_{3} = \qty{-180}{\degree}$, the third harmonic is of same sign with respect to the first harmonic, when the field is maximum. This corresponds to a divergent-like response (Fig.~\ref{subfig:phasor -180}). This response is found at low field for almost all ferroelectrics and relaxors.
    \item $\delta_{3} = \qty{-270}{\degree}$, the third harmonic constricts the center of the loop. This corresponds to the pinching of the loop (Fig.~\ref{subfig:phasor -270}). Pinching of the loop can occur for fresh state for hard ferroelectrics~\cite{MorozovJAP2008} or relaxor above global phase transition temperature~\cite{NadaudAPL2024}.
\end{enumerate}

For relaxors, the non linearities are also studied but using the susceptibility decomposition of the polarization:
\begin{equation}
    P = \varepsilon_{0}\left(\chi_{1}E+ \chi_{2}E^{2}+ \chi_{3}E^{3} + \dots\right)
\end{equation}
where $\chi_{i}$ is the $i$\textsuperscript{th} order susceptibility.

The third order susceptibility is widely studied but often it is only its real part~\cite{DecFerroelectrics2008,MigaPT2006,MigaRSI2007,MigaF2008,KleemannAPL2013}, since for many models dealing with relaxors, the imaginary part is null~\cite{ShettyJAP2020}.
The sign of this contribution can be used to determine the framework explaining the PNR contribution to the non linearities~\cite{PircPRB2002,PircJAP2012,DecFerroelectrics2008,PircPRB1999}.

$\chi_{i}$ can be obtained from the measurement, by applying a sinusoidal electric field, and are a linear combination of $\varepsilon_{r}^{(n)}$, with $n\geq i $~\cite{MigaRSI2007,ShettyJAP2020}.
For example, we have for the third order susceptibility:~\cite{ShettyJAP2020,MigaRSI2007}
\begin{equation}
    \chi_{3} = -\frac{4}{E_{0}^{2}}\left(\varepsilon_{r}^{(3)} + 5\varepsilon_{r}^{(5)} + 14\varepsilon_{r}^{(7)}\right)
\end{equation}
Even if the consideration of the higher order harmonics allows gaining a higher precision on the extraction of the $\chi_{3}$ value ~\cite{MigaRSI2007}, it can lead to a significant increase of the noise ~\cite{DecFerroelectrics2008}.
It is thus possible to consider as a first approximation (especially for weak non-linearity at low driving field $E_{0}$):
\begin{equation}
    \chi_{3} = -\frac{4}{E_{0}^{2}}\varepsilon_{r}^{(3)}
\end{equation}
The dynamics of the PNRs can be thus obtained by studying the sign of $\varepsilon_{r}'^{(3)}$, more specifically if $\delta_{3}$ is close to \qty{0}{\degree} or \qty{-180}{\degree}.
In relaxor ferroelectrics, the total dielectric response is the sum of different contributions (PNRs, ferroelectric domain walls, polar boundaries, etc.) and depending on the measurement conditions, frequency or magnitude of the AC field, temperature, their respective contribution may change~\cite{ShettyJAP2020}.
The decorrelation of the respective contributions is thus made by measuring the dielectric properties at different magnitude and frequency of the measuring AC field.

In addition to impedance spectroscopy, First Order Reversal Curve (FORC) measurements, which allow the extraction of the Preisach distribution~\cite{FujiiUFFC2010,MitoseriuPAC2009,RicinschiJOAM2004} can be used.
Preisach model has been initially developed for the description of the ferromagnetic hysteresis loop ~\cite{PikePRB2003,HarrisonGGG2008} and has been extended to ferroelectric materials ~\cite{MitoseriuJECS2007,FujiiUFFC2010,MitoseriuPAC2009,RicinschiJOAM2004}.
In this model, the $P(E)$ hysteresis loop can be seen as the sum of a large number of elementary hysteresis loop called hysterons.
Each hysteron has given values of of up-switching field $E$ and down switching $E_{r}$.
The knowledge of the hysteron decomposition of the $P(E)$ cycle gives information on the fatigue and polarization of the material ~\cite{ZhuJAP2011,FujiiUFFC2010,StoleriuPRB2006,HoffmannAPL2017}.
It also allows the decomposition of a material in different phases ~\cite{RobertsJGR2006,MuxworthyJGR2005,SaremiPRM2018,NadaudAPL2021}.
Depending on the shape of the hysteresis loop, peculiar FORC distribution can be found~\cite{RobertAPL2000}.

In this article we study the effect of the frequency of the AC measuring signal and the number of measurement cycles on the dielectric response in BCTZ relaxor thin film.
The objective is to show that the pinching response of the studied material is strongly correlated with the presence of PNRs.
In addition, we show that the pinching, and thus the presence of PNR, is strongly influenced by the sub-coercive applied electric field used for the measurement.
FORC measurements are also used to study the pinching response and a novel unipolar impedance measurement method is proposed to probe specific portions of the Preisach plane.

\section{Experiments}
Polycristalline \qty{380}{nm} thick BCTZ film has been grown on \ce{Pt/TiO2/MgO} substrate by pulsed laser deposition. 
Details on growth conditions can be found elsewhere\cite{NadaudAPL2024, NadaudAEM2024}.  
Top circular Au/Ti electrodes (\qty{150}{\um} radius) were deposited through a shadow mask. 
The metal-insulator-metal topology has been chosen for the simple extraction of the dielectric properties using the parallel plate capacitor formula.
This composition is close to the morphotropic phase boundary, and a conventional ferroelectric behavior is found when the material is in ceramic form\cite{LiuPRL2009,KeebleAPL2013}.
In the present case, a relaxor behavior has been observed (see \cref{supfig:realPerm FREQ TEMP} in Supplemantary material \cite{suppMat}, see also references \cite{MaAPL2013,DamjanovicAPL2012,FengCI2019,GaoPSSA2020,LiuCI2023,XuJACS2015,BjrnetunHaugenJAP2013,BenabdallahJAP2011,XuJACS2015,BhardwajJPCS2013,LinAPA2012,LiAFM2018,BokovJMS2006} therein).

The dielectric characterizations presented in this article have been acquired using a lock-in amplifier (MFLI with MD option, Zurich Instrument) connected to a probe station (Summit 12000, Cascade Microtech).
For the impedance and harmonics analysis, the AC measuring signal has been generated using the embedded lock-in amplifier generator.
Its amplitude has been swept from \qty{10}{mV_{rms}} to \qty{1}{V_{rms}} at a frequency from \qty{31}{Hz} to \qty{100}{kHz}.
The applied voltage and current through the capacitor are measured by the lock-in amplifier and are demodulated simultaneously.
The capacitance and the harmonics measurements are described elsewhere~\cite{NadaudJALCOM2022,NadaudAPL2021,NadaudJAP2023,NadaudAPL2024}.
Dielectric permittivity is obtained using parallel plate formula.
Since measurement of the third-harmonic phase-angle can be tricky, careful setting of the lock-in amplifier has been made.
Filters with a order of 8 have been selected.
The bandwidth of the low-pass filter has been set to have a minimum of \qty{80}{dB} of rejection at the demodulated frequency, corresponding to factor 10 between the excitation frequency and the \qty{-3}{dB} cut-off frequency of the filter.
To further increase the sensitivity of the measurement a minimum of 15$\tau$/\qty{0.1}{s} for the averaging is considered.

FORC measurement have been acquired using the same lock-in amplifier, but with a demodulation frequency of \qty{0}{Hz}, which allows having the time evolution of the applied voltage and the resulting current.
The applied voltage has been generated using an Agilent 33250A Waveform Generator.
Numerical integration has been done to obtain the polarization.
To avoid poling effects inevitable with conventional FORC measurement, starting from a saturation state~\cite{FujiiUFFC2010,MitoseriuPAC2009,RicinschiJOAM2004}, we use a bipolar waveform~\cite{CimaRSI2002,NadaudAPL2021,NadaudTSF2023}, allowing to access of the FORC distribution of the fresh state.
This method is then called symmetric FORC.
The detailed procedure of the FORC distribution extraction is given in refs.~\cite{NadaudAPL2021,CimaRSI2002}.
For the FORC distributions, $E$ is the switching field for increasing field and $E_{r}$ is the switching field for the decreasing one.
Unless otherwise specified, all the characterization have been conducted at \qty{300}{K}.

\section{Results and discussion}
The first part is dedicated to measurements as function of frequency and allows us to identify the frequency dependency of the pinching behavior.
The second part contains the study as function of the number of measurements, and thus its effect on the non-linear response.
In the third part, FORC measurements are used to gain a deeper insight into the pinching response.
Finally, unipolar impedance measurement is introduced as a novel method to to probe only a desired part of the FORC distribution at low field.

\subsection{Influence of measurement frequency}
\begin{figure*}
    \centering
    {\includegraphics[width=\textwidth]{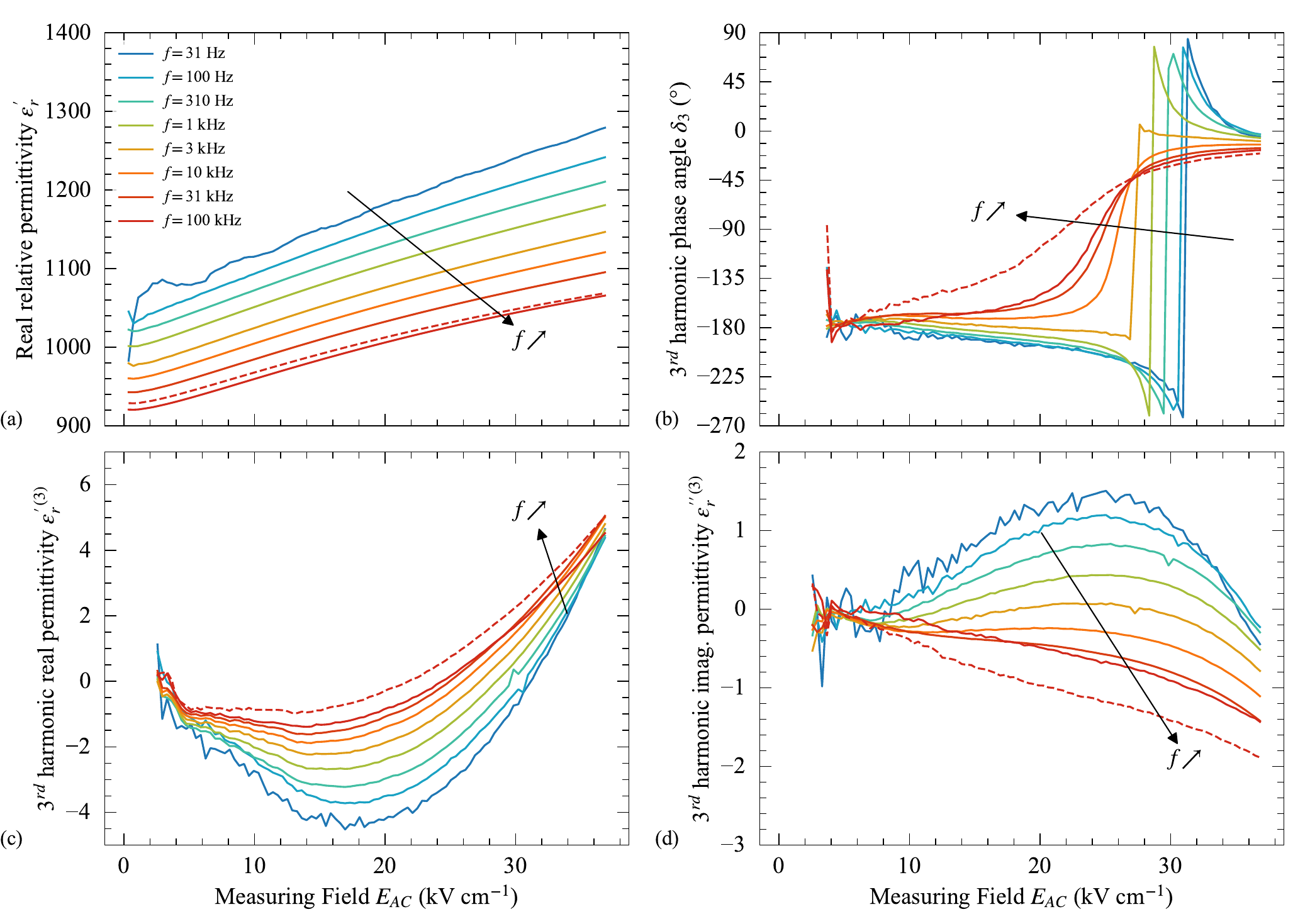}
    \subfloat{\label{subfig:realPerm FREQ OLEV}}%
    \subfloat{\label{subfig:delta3 FREQ OLEV}}%
    \subfloat{\label{subfig:realPerm3 FREQ OLEV}}%
    \subfloat{\label{subfig:imagPerm3 FREQ OLEV}}}%
    \caption{Real part of the first harmonic contribution to the relative permittivity (a), phase-angle (b), real part (c) and imaginary part (d) of the third harmonic contribution as a function of the AC measuring field. Measurement has been made by decreasing the frequency, i.e. from \qty{100}{kHz} to \qty{31}{Hz}, to avoid misinterpretation due to cycling effect discussed afterward. Dashed line corresponds to the measurement at \qty{100}{kHz} made after all the other measurements, to verify that the evolution of the shape of the curve is not related to cycling but a true frequency effect.}
    \label{fig:realPem modulPerm3 delta3 FREQ OLEV}
\end{figure*}

\cref{fig:realPem modulPerm3 delta3 FREQ OLEV} shows the first and third harmonics contribution to the permittivity, as a function of the applied AC electric field, for different frequencies.
Measurement is slightly noisier for low frequencies, which is common~\cite{DecFerroelectrics2008,KleemannAPL2013,NadaudAPL2024}.
This comes from the capacitive behavior of the measured device, which implies a decreasing current value when the measurement frequency decreases.
When the measuring field increases, the relative permittivity (\cref{subfig:realPerm FREQ OLEV}) increases, corresponding to irreversible contribution from the motion of polar boundaries (domain walls pinning/unpinning, polar cluster boundaries, or phase boundaries)~\cite{HashemizadehAPL2017,NadaudAPL2024,damjanovicrpp1998}.
When the frequency change, a very similar shape of the curve is observed.
The only difference consists in a decrease of the permittivity, when the frequency increases, the slope versus electric field staying constant.

The influence of the frequency on the phase-angle of the third harmonic is visible in \cref{subfig:delta3 FREQ OLEV}.
For frequency between \qty{10}{kHz} and \qty{100}{kHz}, $\delta_{3}$ goes from \qty{-180}{\degree}, divergent like contribution, to \qty{-20}{\degree}, saturation like contribution, and passing by \qty{-90}{\degree}.
This kind of response is common for relaxor material ~\cite{NadaudAPL2024,OtonicarJACS2022,HashemizadehAPL2017,OtonicarAFM2020}.

For a frequency of \qty{3}{kHz}, the limit values at low and high fields are the same, but the transition is very steep.
For frequency below \qty{1}{kHz}, the shape of the curve is different.
Instead of increasing after \qty{8}{\kV\per\cm}, the phase-angle decreases and does not pass through \qty{-90}{\degree} but to \qty{-270}{\degree}/\qty{+90}{\degree} which is characteristic of a pinching of the hysteresis loop~\cite{MorozovJAP2008}.

The pinching of the hysteresis loop can correspond to (i) an electric field-induced transition from a paraelectric to a ferroelectric phase, as in \ce{BaTiO3}~\cite{MerzPR1953}, \ce{$(1-x)$Bi_{1/2}Na_{1/2}TiO3-$x$BaTiO3} ~\cite{SapperJAP2014}, (ii) a fresh state in \ce{PbZr_{0.58}Ti_{0.42}O3}~\cite{MorozovJAP2008} or (iii) an aged state and oxygen vacancies migration in Cu-doped in \ce{BaTiO3}~\cite{LiJECS2022}, Ce-doped \ce{Ba(Ti_{0.99}Mn_{0.01})O3}~\cite{ZhaoCI2017}.
Here, the AC field is low and thus the sample aging and charge migration are not expected.
Another possibility is the presence of PNRs that can be oriented by the application of an AC electric field during the measurement~\cite{NadaudAPL2024}.
In the present case, the pinching is well visible even below the transition temperature ($T_{m} = \qty{330}{K}$ here) and is more visible for low frequencies.
This frequency sensitivity supports the idea that the observed pinching signature ($\delta_{3} \simeq \qty{-270}{\degree}$ or \qty{+90}{\degree}) may be related to the PNRs since their response is very frequency dependent~\cite{DecFerroelectrics2008,GlazounovPRB2000,DecPRB2003}.

To verify that the evolution of the $\delta_{3}$ curve is not related to cycling but a true frequency effect, a measurement at \qty{100}{kHz} has been performed after all the measurements presented in \cref{fig:realPem modulPerm3 delta3 FREQ OLEV}.
The shape of the phase-angle 3\textsuperscript{rd} harmonic is globally the same as the one presented in \cref{subfig:delta3 FREQ OLEV}, confirming the frequency influence on the pinching effect.
The difference is attributed to a slight cycling effect, detailed later on.

It is also possible to analyze the third harmonic response by studying it in-phase and out-of-phase, respectively real and imaginary parts (\cref{subfig:realPerm3 FREQ OLEV,subfig:imagPerm3 FREQ OLEV}).
When the AC field increases, the sign of the real part of the third harmonic contribution changes, when the AC field becomes sufficiently high to overcome the quenched state of the PNR\cite{DecFerroelectrics2008,KleemannF2007}.
When the frequency increases, the real part in the range [\qty{3}{\kV\per\cm}, \qty{20}{\kV\per\cm}] significantly decreases in absolute value, diminution of \qty{50}{\%}.
This stronger frequency decay for the third harmonic than for the first one is common for relaxors in which the nonlinearities have for origin PNRs ~\cite{DecFerroelectrics2008,GlazounovPRB2000,DecPRB2003}.

For relaxors in ceramic form, the imaginary part is very small compared to the real part ~\cite{OtonicarAFM2020}.
In the present case the imaginary part is in the same order of magnitude as the real part, which suggests that the total third harmonic response is a sum of different contributions i.e. PNRs and domain boundaries contributions.
For $f = \qty{31}{kHz}$, the imaginary part is negative and its absolute value increases linearly between \qty{4}{\kV\per\cm} and \qty{14}{\kV\per\cm}, which corresponds to conventional Rayleigh response (i.e irreversible domain wall motion contribution).
The almost constant value and close to zero for fields below \qty{4}{\kV\per\cm} corresponds to the low field region in which the polar boundaries contribution is mostly reversible~\cite{SchenkPRA2018,BassiriGharbJE2007}.
Above \qty{24}{\kV\per\cm}, the imaginary part decreases, which reveals that the domain wall motion contribution no longer follows the Rayleigh law, in agreement with the saturation observed on the first harmonic permittivity (\cref{subfig:realPerm FREQ OLEV}).
When the frequency decreases, the imaginary part at AC fields higher than \qty{25}{\kV\per\cm} progressively goes from negative values to positive values.
If only domain wall motion would be present, the imaginary part should stay negative.
The progressive change of sign indicates that another contribution to the permittivity (PNRs) progressively takes place and induces the pinching visible for low frequencies.
For low AC field (below \qty{10}{\kV\per\cm}), the evolution with frequency is almost not visible.

Fitting of the $\varepsilon_{r}'(E_{\mathit{AC}})$ curves (\cref{subfig:realPerm FREQ OLEV}) allows the extraction of the polar boundaries contributions to the permittivity as function of the frequency, reported in Supplementary Material \cite{suppMat} (see also references \cite{GharbJAP2005,BassiriGharbJE2007,BassiriGharbJAP2006,BeckerJAP2022,CollaJAP1999,DamjanovicMSE2005, jonscherjopd,NadaudJAP2015,CoulibalyAPL2020,jonscherbook} therein and \cref{supfig:perm coeff FREQ OLEV}).

\begin{figure*}
    \centering
    {\includegraphics[width=\textwidth]{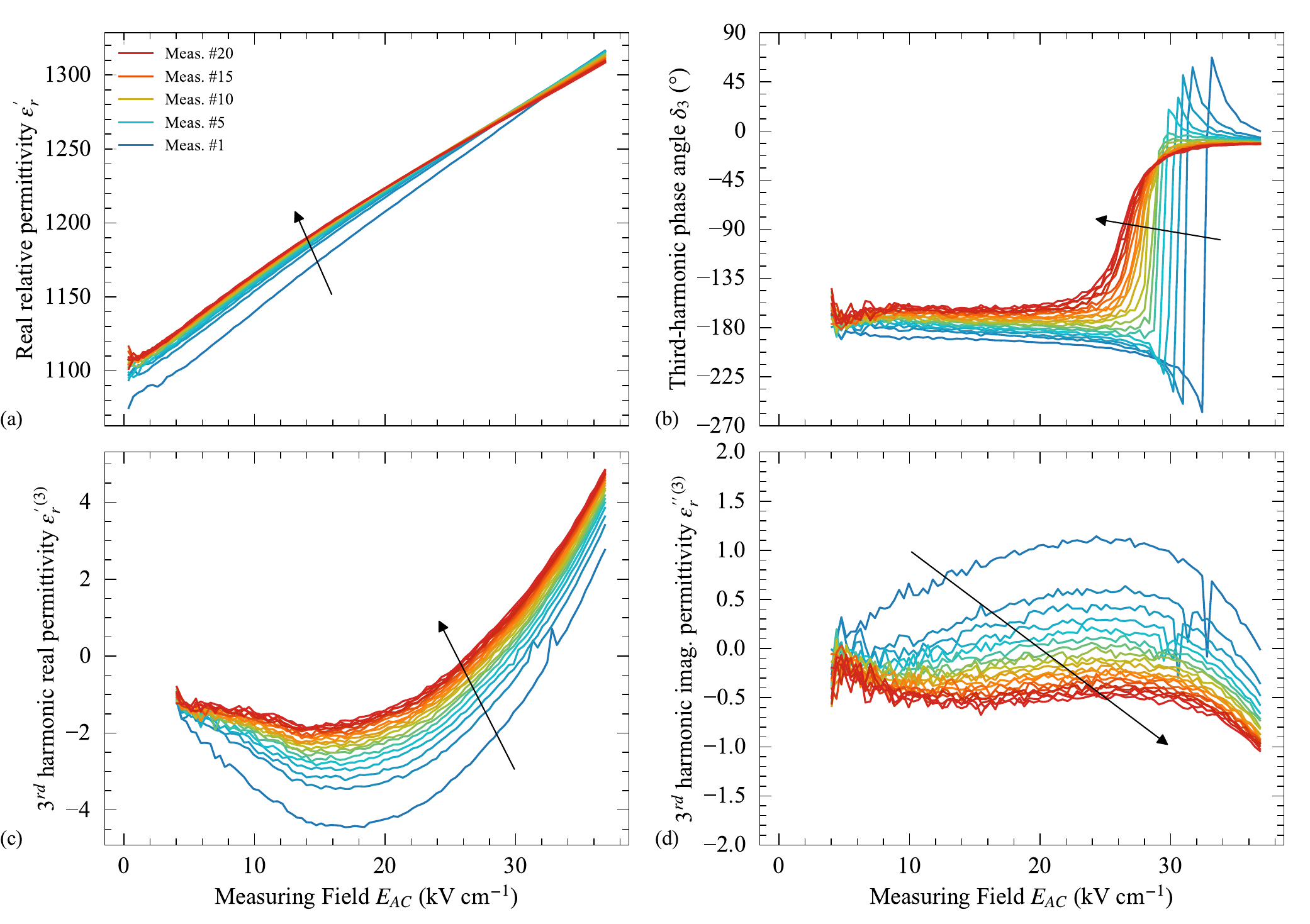}
    \subfloat{\label{subfig:realPerm NUM OLEV}}%
    \subfloat{\label{subfig:delta3 NUM OLEV}}%
    \subfloat{\label{subfig:realPerm3 NUM OLEV}}%
    \subfloat{\label{subfig:imagPerm3 NUM OLEV}}}%
    \caption{Real part of the permittivity (a), the phase-angle (b), the real part (c) and the imaginary part (d) of the third harmonics, as a function of the measurement electric field, for different number of measurements, for $f=\qty{1}{kHz}$.}
    \label{fig:realPerm realPerm3 delta3 imagPerm3 NUM OLEV}
\end{figure*}

\subsection{Influence of the number of measurements (cycling)}

\begin{figure}
    \centering
    {\includegraphics[width=0.45\textwidth]{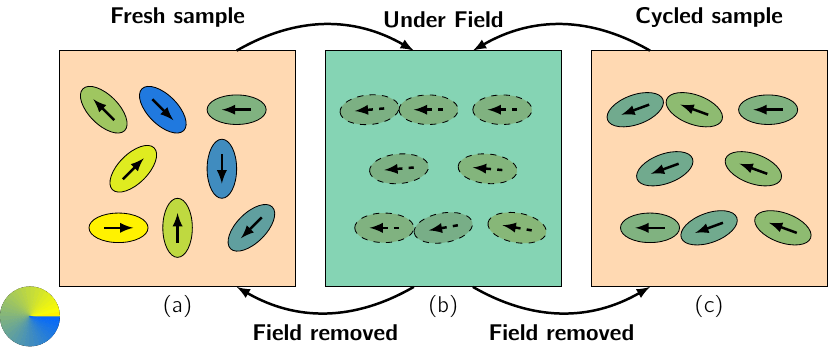}%
    \subfloat{\label{subfig:pinching initial}}%
    \subfloat{\label{subfig:pinching field}}%
    \subfloat{\label{subfig:pinching cycled}}}%
    \caption{Illustration of the PNR configuration for a fresh sample (a), for a sample under field (b) and for a cycled sample (c).}
    \label{fig:pinching sequence}
\end{figure}

\cref{fig:realPerm realPerm3 delta3 imagPerm3 NUM OLEV} shows the first and third harmonics contribution to the permittivity as a function of the measuring electric field, for successive measurements.
The relative permittivity exhibits a very small increase after the first measurement and for the following measurements, the increase is almost imperceptible (\cref{subfig:realPerm NUM OLEV}).
By difference, the third harmonic response shows a clear evolution as a function of cycling.
For the first 5 measurements, the phase-angle of the third harmonic exhibits the pinching response which has been seen in the previous study concerning the frequency effect (\cref{subfig:delta3 NUM OLEV}).
After the \nth{6} measurement, the pinching signature ($\delta _{3} = \qty{-270}{\degree}$) is no more visible and the phase-angle response corresponds more to what is obtained for relaxor ferroelectrics ($\qty{-180}{\degree}\rightarrow \qty{-90}{\degree} \rightarrow \qty{0}{\degree}$).
When increasing the number of measurements, the phase response shape slightly evolves: the transition is less steep and the field at which $\delta_{3}=\qty{-90}{\degree}$ decreases.
The phase-angle of the third-harmonic at low field exhibits a plateau between \qty{4}{\kV\per\cm} and \qty{20}{\kV\per\cm}, which value increases from \qty{-180}{\degree} to \qty{-145}{\degree}.
This later value is close to what is found for the ferroelectric-relaxor (RE-FE) category given in ref~\cite{OtonicarAFM2020}.

The existence of a pinched hysteresis loop upon the first field cycles results from the following dipole orientation sequence.
At zero field, dipoles are mostly randomly oriented (\cref{subfig:pinching initial}), leading to a small but non-zero polarization. 
As the electric field increases in magnitude, dipoles orient in the same direction (\cref{subfig:pinching field}, like for a conventional ferroelectric) and a large polarization is present.
When the field magnitude is decreased, large backswitching occurs (\cref{subfig:pinching initial}) and contrary to conventional ferroelectric, a strong decrease of the polarization is visible.
This leads to low value of the polarization around zero field, and thus the pinching/constricted loop~\cite{ZhaoCI2017,MorozovJAP2008}.
However, when repeating the measurement multiple times, less and less backswitching occurs. 
Most PNRs remain oriented in the same direction and thus applying again the AC field does not lead to a large re-orientation  i.e. the pinching is fading upon cycling (\cref{subfig:pinching cycled}).

The effect of the number of measurements performed on the real and imaginary parts of the third harmonic contribution, visible in \cref{subfig:realPerm3 NUM OLEV,subfig:imagPerm3 NUM OLEV}, is very similar to the frequency effect (\cref{subfig:realPerm3 FREQ OLEV,subfig:imagPerm3 FREQ OLEV}).
When repeating the measurement, the real part at low AC field decreases, in absolute value, and the field for which the sign changes is lower.
This confirms that after multiple measurements, the applied AC electric field leads to a lower response since PNRs have been oriented in the same direction. 
The value at high AC field slightly increases, corresponding again to a saturation, ($\delta_{3}\simeq \qty{0}{\degree}$) of the permittivity, which is more visible on the first harmonic after a large number of measurement (\cref{subfig:realPerm NUM OLEV}).
The evolution of the pinching when measuring multiple times the material indicates that even at fields lower than the coercive field the domain/phase structure can be influenced.

Fitting of the $\varepsilon_{r}'(E_{\mathit{AC}})$ curves (\cref{subfig:realPerm NUM OLEV}) allows the extraction of the polar boundaries contributions to the permittivity as function of the cycle number, reported in Supplementary information \cite{suppMat} (see \cref{supfig:coeff NUM OLEV} and refs. \cite{QiuJAP2019,HeJJAP2019,TianAPL2023,borderonapl2011,BorderonSR2017,GharbJAP2005} therein).

\subsection{First Order Reversal Curves and unipolar Rayleigh measurements}

\begin{figure*}
    \centering
    \begin{minipage}[c]{0.63\textwidth}
    {\includegraphics[width=\textwidth]{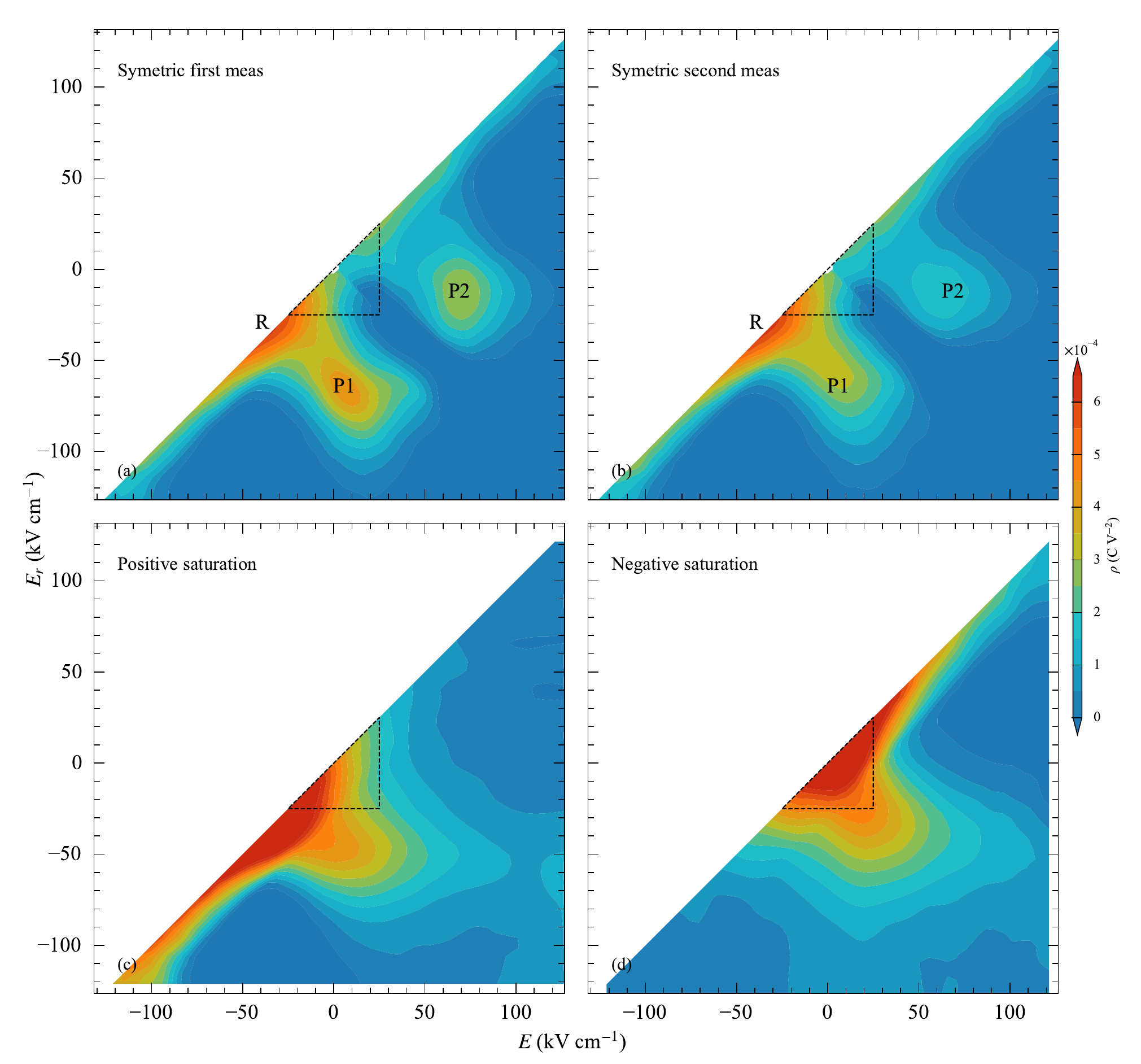}%
    \subfloat{\label{subfig:FORC first}}%
    \subfloat{\label{subfig:FORC second}}%
    \subfloat{\label{subfig:FORC sym pos}}%
    \subfloat{\label{subfig:FORC sym neg}}%
    }%
    \end{minipage}
    \begin{minipage}[c]{0.35\textwidth}
    {\includegraphics[width=\linewidth]{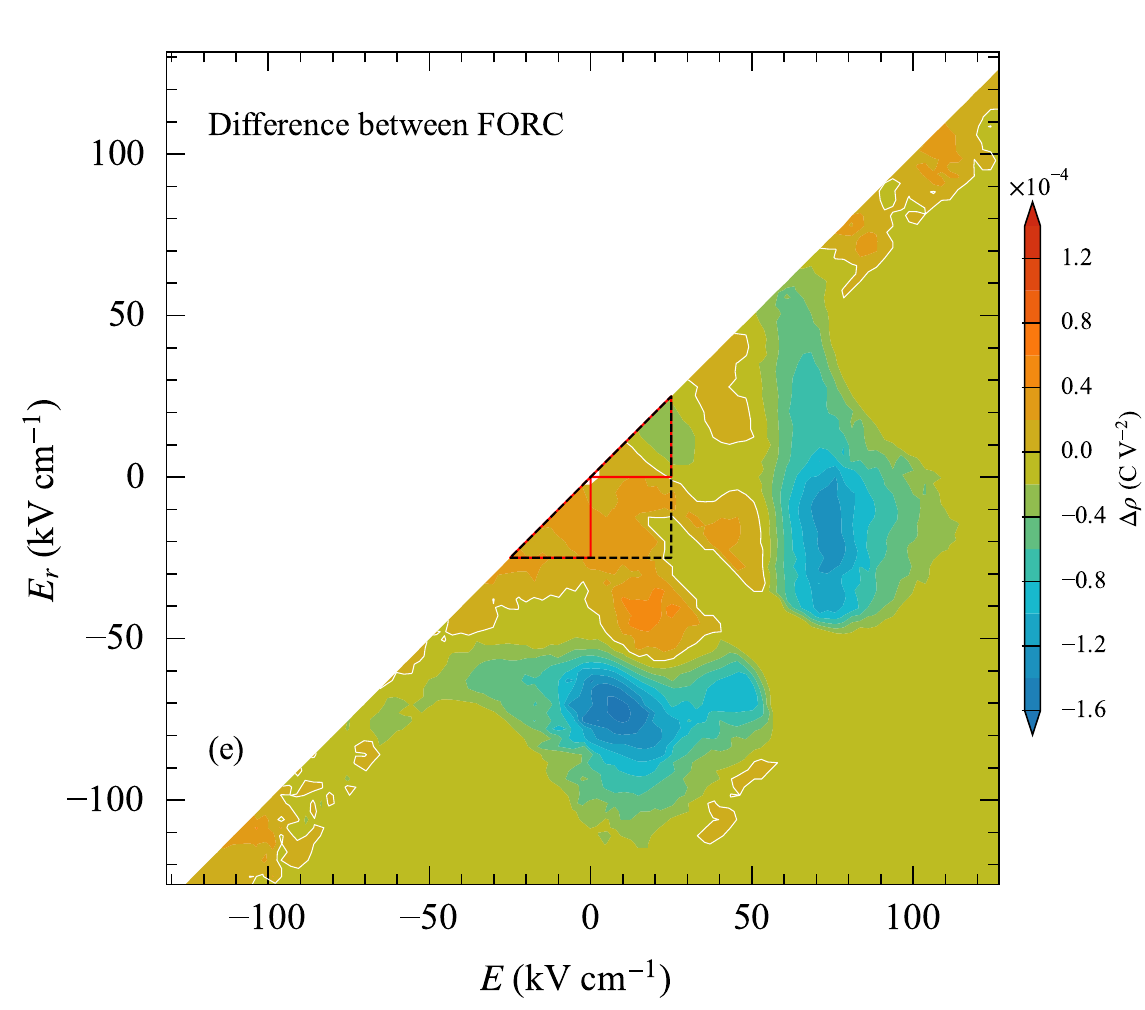}
    \subfloat{\label{subfig:diff FORC}}}%
    \subfloat[\label{tab:weigth table}]{%
    {\resizebox{\linewidth}{!}{%
    \begin{tabular}{p{0.4\linewidth}p{0.45\linewidth}p{0.45\linewidth}p{0.45\linewidth}}
        \hline
        \hline
        &\multicolumn{3}{c}{Contribution} \tabularnewline
        \cline{2-4}
        Measurement & R     & P1   & P2   \tabularnewline
        \hline
        First     & \qty{15.39}{\micro\coulomb\per\cm\squared} & \qty{1.18}{\micro\coulomb\per\cm\squared} & \qty{0.66}{\micro\coulomb\per\cm\squared} \tabularnewline
        Second    & \qty{15.47}{\micro\coulomb\per\cm\squared} & \qty{0.77}{\micro\coulomb\per\cm\squared} & \qty{0.19}{\micro\coulomb\per\cm\squared} \tabularnewline
        \hline
        \hline
    \end{tabular}}%
    }}
    \end{minipage}
    \caption{FORC distribution for the first (a) and second (b) symmetric measurements and for positive saturation (c) and negative saturation (d) measurements. Difference between the symmetric measurements (e). Solid white line in (e) indicates the frontier between positive and negative values. Black dashed lines corresponds to the area which is probe for symmetric AC measurements. Red lines in (e) indicates the area probe for unipolar impedance measurement.%
        (f) Contribution to the polarization of the reversible contribution (R) and the two peaks (P1) and (P2).}

    \label{fig:FORC}
\end{figure*}

FORC analysis has been performed to estimate local internal field $E_{i}$ / coercive field $E_{c}$ distribution and understand the effect of successive measurements on the material response.
$P(E)$/$I(E)$ loops used for the FORC extraction are visible in Supplementary material \cite{suppMat} (see \cref{supfig:polar,supfig:current} and referencesc \cite{NadaudAPL2024,CaoJACS2019,LiJECS2022,HashemizadehAPL2017} therein.).
The distribution given by first symmetric FORC measurement is presented in \cref{subfig:FORC first}.
Close to the axis $E = E_{r}$, at $(E, E_{r}) = (\qty{-30}{\kV\per\cm}, \qty{-30}{\kV\per\cm})$ (noted R), a peak is visible corresponding to a large reversible contribution, typical from soft ferroelectric/relaxors~\cite{FujiiJACS2011}.
Moreover, it exhibits two peaks (noted P1 and P2), similarly to what is observed for unpoled PZT~\cite{CimaRSI2002} or pristine state \ce{Hf_{0.5}Zr_{0.5}O2}~\cite{FenglerAEM2017,FenglerJAP2018,SchenkAMI2015}, and these two peaks are typical of the pinched hysteresis loop ~\cite{RobertAPL2000}.
The peak positions are almost symmetric $(E, E_{r}) = (\qty{69}{\kV\per\cm}, \qty{-17}{\kV\per\cm})$ and $(\qty{6.2}{\kV\per\cm}, \qty{-66}{\kV\per\cm})$.
These two peaks in the FORC distribution corresponds to the current peaks visible in \cref{supsubfig:current meas 1} at low field.
The peak separation is not visible for large field measurements due to the orientation of the PNR, similar to what is found for the cycling study \cref{fig:pinching sequence,fig:realPerm realPerm3 delta3 imagPerm3 NUM OLEV}.

When repeating the symmetric FORC measurement a second time (\cref{subfig:FORC second}), the distribution still presents two peaks at the same positions but their intensity is greatly reduced, especially the peak in the upper right corner (at \qty{6.2}{\kV\per\cm}, \qty{-66}{\kV\per\cm}). 
The hysteron decomposition obtained for the second measurement is thus closer to what is expected for a soft ferroelectric~\cite{FujiiJACS2011}, a maximum of hysteron density close to the axis $E_{r} = E$.
\cref{subfig:diff FORC} shows the difference between the first and the second symmetric FORC measurements.
In addition to the large decrease of the main peaks, the hysteron density close to the origin increases.

To quantify the contribution of each peak and of the reversible part to the total polarization, FORC distribution has been integrated using the following expression:\cite{FujiiUFFC2010}
\begin{equation}
    P = \iint p(E, E_{r})d E d E_{r} + \int p_{\mathit{rev}}(E)d E
\end{equation}
where $p(E, E_{r})$ and $p_{\mathit{rev}}(E)$ are the Preisach distribution for irreversible ($E>E_{r}$) and reversible hysterons ($E=E_{r}$), respectively.
The results of the integration are given in Table~\ref{tab:weigth table}.
For the first and second measurements, the reversible contribution is 8.4 and 16 times larger than the sum of the 2 peaks related to PNR's, respectively.
This explains the reason why the pinching is not visible on the $P(E)$ loops (\cref{supfig:polar}), enhancing the interest of the third harmonic measurement and the FORC to reveal a pinching contribution to the polarization.

In addition, FORC measurements have been done by starting from a saturated state (\cref{subfig:FORC sym pos,subfig:FORC sym neg}).
In both cases, the distribution presents only one peak, very close to the $E_{r} = E$ axis, which corresponds to what is expected for a soft ferroelectric~\cite{FujiiJACS2011}.
Depending on the initial state (positive or negative saturation), a slight shift of the peak is visible.
This peak shift indicates an imprint is present~\cite{ChojeckiJAP2021}, which may arise from a different domain structure for positive and negative bias field.
The peak position is very close to the one subsists while doing bipolar FORC measurement.
The presence of only one peak shows that the application of a large electric field remove the pinching response of the material and the response correspond to a conventional soft ferroelectric.

A third symmetric FORC measurement has been made after the two starting from saturation.
No significant changes compared to the second symmetric FORC is visible, the distribution still exhibits two peaks, but with a magnitude lower than for the first measurement (given in supplementary material \cite{suppMat}, \cref{supfig:forc2 forc3}).

\begin{figure}[h!]
    \centering
    \subfloat[\label{subfig:preisach AC}]{\includegraphics[width=0.23\textwidth]{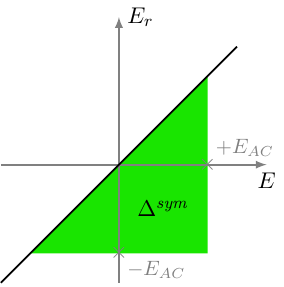}}%
    \subfloat[\label{subfig:preisach uni}]{\includegraphics[width=0.23\textwidth]{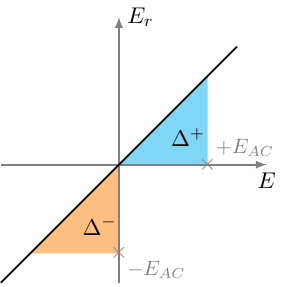}}%

    {\includegraphics[width=0.48\textwidth]{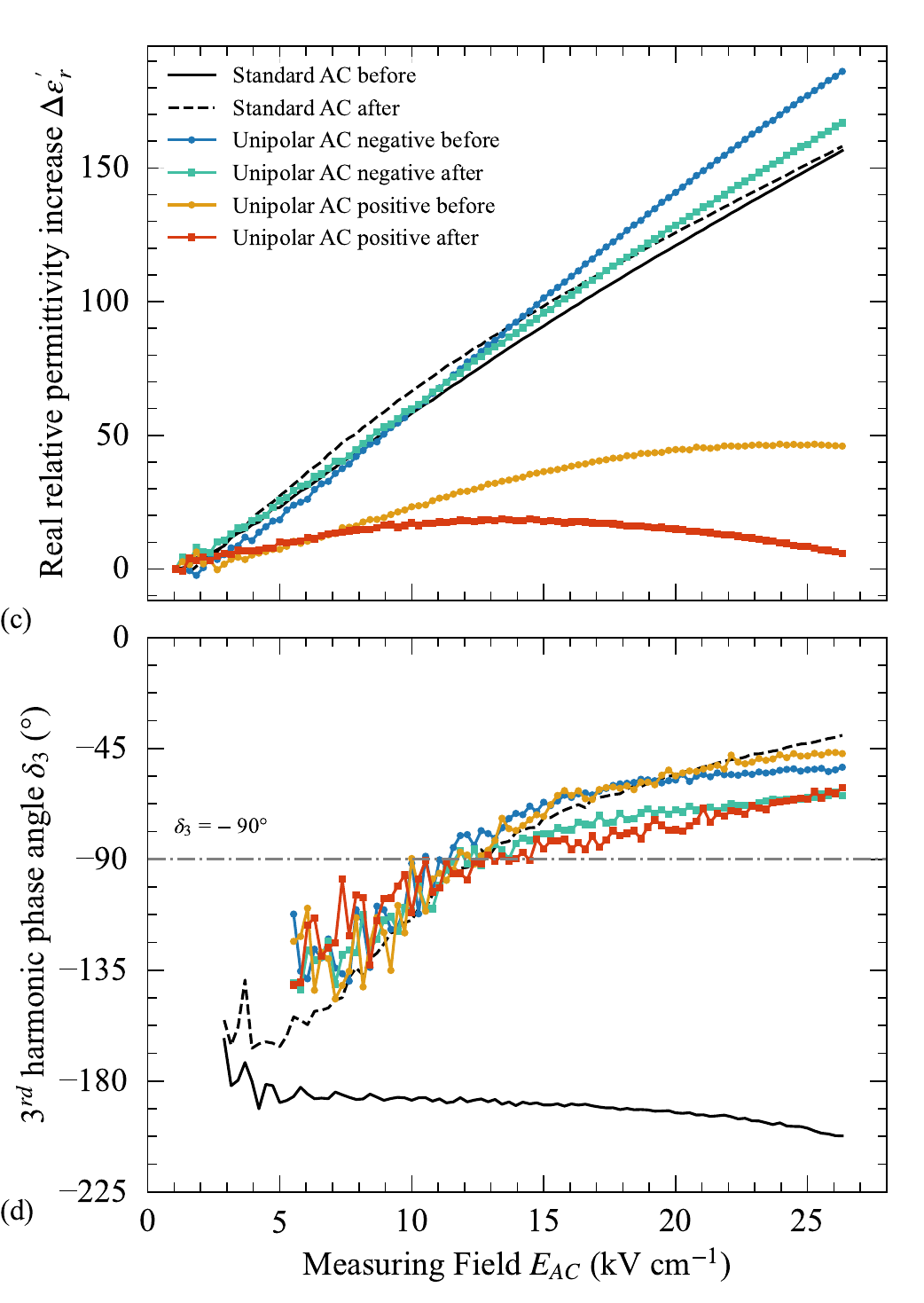}%
    \subfloat{\label{subfig:unipolar permittivity}}%
    \subfloat{\label{subfig:unipolar delta3}}}%

    \caption{Preisach plane areas probed when doing conventional AC measurement (a) and using unipolar AC signal (b). Cyan area corresponds to the portion of the plane which is probed using positive signal and orange to the negative. Real part of the permittivity (c) and phase-angle of the third harmonic (d) measured using the unipolar AC signal given \cref{eq:unipolar signal} (colored curves) and using conventional impedance spectroscopy (black lines). Gray dash-dotted line in (d) corresponds to phase angle for a homogeneous hysteron density.}
    \label{fig:unipolar permittivity}
\end{figure}

For an  ideal material following the Rayleigh law, a homogeneous distribution of hysteron is present at low fields~\cite{RobertJAP2001}.
When doing conventional AC measurement, the part of the FORC distribution that is probed is symmetric according to the axis $E_{r} = -E$, and corresponds to the triangle $\Delta^{\mathit{sym}}$ defined by the vertices $(-E_{\mathit{AC}}; -E_{\mathit{AC}})$ , $(+E_{\mathit{AC}}; -E_{\mathit{AC}})$ and $(+E_{\mathit{AC}}; +E_{\mathit{AC}})$ as shown \cref{subfig:preisach AC}.
The increase of the permittivity assuming a homogeneous irreversible hysterons distribution may be calculated:~\cite{FujiiUFFC2010,FujiiJACS2011,RobertJAP2001}
\begin{equation}
    \Delta \varepsilon_{r} = \frac{\rho A^{sym}}{2\varepsilon_{0} E_{\mathit{AC}}}
\end{equation}
$A^{sym}$ corresponds to the area of the triangle $\Delta^{\mathit{sym}}$ and $\rho$ to the hysteron density in the homogeneous part.
This gives the linear increases of the permittivity expected by the Rayleigh law:
\begin{equation}
    \label{eq:bipolar rayleigh}
    \Delta \varepsilon_{r} = \frac{\rho E_{\mathit{AC}}}{\varepsilon_{0}} = \alpha^{\mathit{sym}} E_{\mathit{AC}}
\end{equation}
$\alpha^{\mathit{sym}}$ refers to the Rayleigh coefficient in the case of a symmetrical AC measurement.

To probe the asymmetry of the FORC distribution, we propose to use instead a unipolar waveform for the applied electric field which consists in:
\begin{equation}
    \label{eq:unipolar signal}
    E(t) = \pm \frac{E_{\mathit{AC}}}{2}\left(1+\sin\left(2\pi f t + \phi\right)\right)
\end{equation}
The sign $\pm$ denotes if we want to probe the positive hysterons or the negative hysterons (\cref{subfig:preisach uni}).
This signal has the properties to be always positive (or negative) and has a maximum magnitude of $E_{\mathit{AC}}$.
It can be generated directly using the lock-in amplifier by adding to the generated signal the demodulated magnitude.
Permittivity is obtained by using the same formula as conventional AC measurement.
One can note the $\frac{1}{2}$ factor to obtain the same peak value of the signal.

Probing only a part of the Preisach plane is possible using a recently proposed extension of the FORC, namely the Unipolar Reversal Curve (URC)~\cite{VecchiAEM2024}.
Fundamental differences between URC and the proposed unipolar impedance measurement exist since they are complementary.
URC is a large signal method, and thus intends to explore the switching dynamic of the material.
The proposed unipolar impedance measurement is a small signal method and thus aims to explore non switching dynamic of the material.

Since the proposed method is based on impedance measurement, long integration time can be used to increase the signal-to-noise ratio, allowing high sensitivity of the measurement and higher order harmonics are also extracted (magnitude and phase).
This averaging is not straightforward for URC due its large signal nature which can induce fatigue of the sample.
Proposed unipolar impedance measurement should not induce large fatigue effects, due to its small signal nature.
Nevertheless, for the proposed method, the excitation signal needs to be sinusoidal, contrary to URC where delay between excitation pulses may be added, allowing to studied time dependant imprint effect.

The portion of hysteron which are switched correspond to $\Delta^{+}$ and $\Delta^{-}$ for positive and negative unipolar signal, respectively.
Due to symmetry considerations, they are the same and can be calculate using the following formulas:
\begin{equation}
    \Delta \varepsilon_{r}^{-} = \frac{\rho A^{-}}{2\varepsilon_{0} \frac{E_{\mathit{AC}}}{2}} \qquad\text{and}\qquad \Delta \varepsilon_{r}^{+} = \frac{\rho A^{+}}{2\varepsilon_{0} \frac{E_{\mathit{AC}}}{2}}
\end{equation}
Which gives:
\begin{equation}
    \label{eq:unipolar rayleigh}
    \Delta \varepsilon_{r}^{+} = \frac{\rho E_{\mathit{AC}}}{2\varepsilon_{0}} = \alpha^{+} E_{\mathit{AC}}  \quad\text{and}\quad \Delta \varepsilon_{r}^{-} = \frac{\rho E_{\mathit{AC}}}{2\varepsilon_{0}} = \alpha^{-} E_{\mathit{AC}}
\end{equation}
$\alpha^{+}$ and $\alpha^{-}$ refer to the Rayleigh coefficients in the case of a positive and negative unipolar measurement, respectively.
With this new methodology, it would possible to apply the Rayleigh law with only a positive (or negative) signal.
In the case of a homogeneous distribution of hysteron, the Rayleigh coefficient for the unipolar signal is the half of the one for bipolar signal, i.e. $\alpha^{+}=\alpha^{-} = \frac{\alpha^{\mathit{sym}}}{2}$.

\cref{subfig:unipolar permittivity} shows the real part of the permittivity measured using the unipolar AC signal, before and after FORC measurement.
For all the measurements, the permittivity increases when the AC field increases, similarly to what is obtained for conventional AC measurement.
This increase is due to the irreversible polar boundaries contribution and corresponds to the (almost) homogeneous hysteron distribution close to the origin of the Preisach plane.

Rayleigh coefficients have been extracted for all the measurement conditions and are reported in \cref{table:Rayleigh}.
The difference between positive and negative unipolar measurements indicates that the hysteron density is not the same for the positive and negative parts, which is consistent with the FORC measurement presented in \cref{subfig:FORC first}.
According to \cref{eq:bipolar rayleigh,eq:unipolar rayleigh} the increases should be two times lower for unipolar measurements than for bipolar measurements.
In the present case, the increase for negative unipolar and bipolar signals are very similar (values of $\alpha$ close to each other), whereas the increase is lower for unipolar positive than for negative.
This reveals that the hysteron density is not homogeneous and the use of both unipolar and bipolar signals permits to measure this.
After the FORC measurement, the irreversible contribution, represented by the $\alpha$ parameter, is slightly higher (for symmetric and negative unipolar measurements), similar with what has been observed for AC-poling~\cite{QiuJAP2019,HeJJAP2019,TianAPL2023}.
The Rayleigh coefficient for positive unipolar measurement decreases, which corresponds well to what is seen on the FORC density difference (\cref{subfig:diff FORC}), a slight decrease of the hysteron density in the triangle $\Delta^{+}$.

The phase-angle of the third harmonic is also measured and is reported \cref{subfig:unipolar delta3}.
Before FORC, the symmetric measurement exhibits the same shape as seen in \cref{subfig:delta3 NUM OLEV} (one can note the pinching is not visible due to lower maximum field to avoid cycling influence).
For all the others conditions the same global evolution is visible.
For low fields (around \qty{5}{\kV\per\cm}), $\delta_{3}$ is between \qty{-180}{\degree} and \qty{-135}{\degree}, corresponding to a mainly reversible contribution, and for high fields (above \qty{20}{\kV\per\cm}), the limit value is between \qty{-90}{\degree} and \qty{0}{\degree}, corresponding to a mix between irreversible domain wall motion contribution and saturation.
The main difference is the limit value which is closer to the ideal value of \qty{-90}{\degree} for unipolar measurement, indicating areas $\Delta^{+}$ and $\Delta^{-}$ are more homogeneous than $\Delta^{\mathit{sym}}$.

\begin{table} 
\caption{Rayleigh coefficient extraction using \cref{hyperbolic}.}\label{table:Rayleigh}
\centering
\resizebox*{0.48\textwidth}{!}{%
\begin{tabular}{cccc}
                \hline
                \hline
    \multirow{2}{*}{Condition} & \multicolumn{3}{c}{Measurement type} \tabularnewline
                \cline{2-4}
                & Conventional AC            & Unipolar negative          & Unipolar positive          \tabularnewline
                \hline
    Before FORC & \qty{7.0(4)}{\cm\per\kV}       & \qty{7.3(1)}{\cm\per\kV} & \qty{5.0(8)}{\cm\per\kV} \tabularnewline
    After FORC  & \qty{8.4(1)}{\cm\per\kV}       & \qty{7.6(1)}{\cm\per\kV} & \qty{3.4(2)}{\cm\per\kV}   \tabularnewline
                \hline
                \hline
\end{tabular}}
\end{table}

\section{Conclusions}
In this article, the dielectric non-linearities in \ce{0.5(Ba_{0.7}Ca_{0.3}TiO_{3})-0.5(BaZr_{0.2}Ti_{0.8}O_{3})} thin film are studied using impedance spectroscopy and harmonic measurements, as a function of the AC measuring field, at different frequencies and upon cycling.
The measurements reveal a phase angle of the third harmonic close to \qty{-270}{\degree}/\qty{+90}{\degree}, corresponding to a pinching of the hysteresis loop, especially for low frequency.
This strong frequency dependency of the pinching response (i.e. $\delta_{3}$ close to \qty{-270}{\degree}/\qty{+90}{\degree}) indicate that the pinching is induced by the presence of polar nanoregions (PNRs), whose responses are known to be strongly frequency-dependent.

When repeating the measurement, the PNR contribution changes as indicated by the evolution of the third harmonic from a pinched to a conventional relaxor response.
This shows that strong changes in the PNR configuration can be induced even for low AC fields.

First Order Reversal Curve (FORC) measurements confirm the presence of a pinched hysteresis loop by the presence of two peaks in the Preisach distribution.
When repeating the FORC measurement a second time, the distribution drastically changes towards to a soft ferroelectric/relaxor.
This Preisach distribution presents an asymmetry in the low field range and we propose a new method to probe this asymmetry, called unipolar impedance measurement.
This method allows to observe only a specific part of the Preisach plane.
We shown the Rayleigh law is still valid even when measuring the dielectric permittivity with only positive (or only negative) electric field.
In addition to the obvious interest of this new methodology from a characterization point of view, numerous applications use unipolar signals, in which case the variation of the dielectric permittivity may need to be taken into account.

\section*{Supplementary Material}
Analysis of the different contributions to the permittivity (lattice, reversible and irreversible polar boundary motion contributions) as a function of the frequency and the measurement number, $P(E)$ and $I(E)$ loops for the FORC extraction and comparison between the second and third FORC measurements are given is supplementary material.

\section*{Data availability}
The data that support the findings of this study are available from the corresponding author upon reasonable request.

\section*{Acknowledgments}
This work has been performed with the means of the CERTeM (microelectronics technological research and development center) of French region Centre Val de Loire.
This work was funded through the project MAPS in the program ARD+ CERTeM 5.0 by the Région Centre Val de Loire co-funded by the European Union (ERC, DYNAMHEAT, N°101077402) and by the French State in the frame of the France 2030 program through the French Research Agency (ANR-22-PEXD-0018). 
Views and opinions expressed are however those of the authors only and do not necessarily reflect those of the European Union or the European Research Council. 
Neither the European Union nor the granting authority can be held responsible for them.

\section*{Conflict of Interest}
The authors declare no competing financial interest.

%

\end{document}